\documentclass{article}
\usepackage[preprint,nonatbib]{nips_2018}
\usepackage[utf8]{inputenc} 
\usepackage[T1]{fontenc}    
\usepackage{hyperref}       
\usepackage{url}            
\usepackage{booktabs}       
\usepackage{amsfonts}       
\usepackage{nicefrac}       
\usepackage{microtype}      
\usepackage{xcolor}
\usepackage{bm}
\usepackage{amsmath}
\usepackage{graphicx}
\usepackage{amsmath}
\usepackage{float}
\usepackage[export]{adjustbox}
\usepackage{subcaption}
\usepackage{makecell}
\usepackage{titling}
\usepackage{tcolorbox}
\usepackage{mathtools}

\definecolor{colour3}{RGB}{178,55,250} 

\newcounter{noteCZctr} 

\newcounter{noteMCctr} 

\newcounter{noteZZctr} 

\DeclarePairedDelimiter\floor{\lfloor}{\rfloor}
\makeatletter

\let\c@table\c@figure
\makeatother

\title{A Universal End-to-End Approach to \\ Portfolio Optimization via Deep Learning}

\newcommand*\samethanks[1][\value{footnote}]{\footnotemark[#1]}

\author{
  Chao Zhang\thanks{Equal contribution. Correspondence
to: Zihao Zhang <zhangzihao@hotmail.co.uk>}, Zihao Zhang\samethanks, Mihai Cucuringu, Stefan Zohren\\
  University of Oxford
}

\begin{document}

\maketitle

\begin{abstract}
 
We propose a universal end-to-end framework for portfolio optimization where asset distributions are directly obtained. The designed framework circumvents the traditional forecasting step and avoids the estimation of the covariance matrix, lifting the bottleneck for generalizing to a large amount of instruments. Our framework has the flexibility of optimizing various objective functions including Sharpe ratio, mean-variance trade-off etc. Further, we allow for short selling and study several constraints attached to objective functions. In particular, we consider cardinality, maximum position for individual instrument and leverage. These constraints are formulated into objective functions by utilizing several neural layers and gradient ascent can be adopted for optimization. To ensure the robustness of our framework, we test our methods on two datasets. Firstly, we look at a synthetic dataset where we demonstrate that weights obtained from our end-to-end approach are better than classical predictive methods. Secondly, we apply our framework on a real-life dataset with historical observations of hundreds of instruments with a testing period of more than 20 years.  

\end{abstract}


\section{Introduction}
\label{introduction}

Portfolio optimization is the process of selecting the best asset distribution to invest in out of all the considered assets, with the aim of optimizing a suitably chosen objective function, such as maximizing returns while minimizing the level of risk. The theory, also known as modern portfolio theory (MPT), was introduced by Markowitz \cite{markowitz1952portfolio} and has become one of the corner stones of quantitative finance. The main benefit of MPT stems from diversification, as the variance of a portfolio decreases with the addition of less correlated instruments, leading to improvements in the portfolio's Sharpe Ratio \cite{sharpe1994sharpe}, one of the popular performance metrics employed by practitioners. Higher Sharpe Ratios can be combined with leverage, leading to higher returns for a given level of risk preference.

Despite the wide popularity and practical usage of MPT, the classical mean-variance (MV) optimization approach employed in MPT severely suffers from a number of limitations. In particular, the traditional method of optimizing a portfolio is a two-step optimization problem, where the first step aims to predict future returns, and a constrained optimization problem is then solved to derive the optimal portfolio weights. However, financial returns are notoriously stochastic with extremely low signal-to-noise ratio. The construction of predictive signals remains extremely difficult, and the generalization and robustness of such signals is often questioned. Furthermore, the MV method involves the estimation of the covariance matrix for all the considered instruments, and computing its inverse \cite{pantaleo2011improved}. However, such estimations are highly unstable, due to the large number of pairwise coefficients to estimate, which 
hinders the scalability of the method and can lead to concentration risk\footnote{Portfolio weights are only assigned to a few instruments, which decreases the diversification power.}.

In this work, we propose an end-to-end framework that directly optimizes a portfolio by utilizing deep learning models \cite{goodfellow2016deep}. Compared to the classical two-step method, we directly optimize portfolio weights circumventing the requirement of predicting future returns. Furthermore, our framework enjoys the flexibility of being able to optimize \emph{various objective functions}, including global minimum variance, maximum Sharpe Ratio, mean-variance trade-off, etc. This flexibility allows us to study portfolios with distinct characteristics, and to adjust risk preference accordingly. Given the chosen objective function, our setup implicitly takes into account the estimation of the covariance matrix, thus  ensuring the scalability of our approach.

Our framework allows for short-selling, and we study \emph{different constraints} attached to objective functions. Specifically, we consider cardinality\footnote{We limit the number of traded assets in a portfolio.}, maximum position\footnote{We limit the maximum position for individual instrument preventing concentration risk.}, short-selling and leverage. The general MV approach is a standard quadratic programming problem,  which can be solved exactly optimally. However, the addition of constraints can increase the degree of complexity; for example, the cardinality constraint renders the problem non-convex, and no longer solvable by exact methods \cite{jin2016constrained}. Our approach allows us to incorporate these constraints into objective functions by utilizing several neural layers, and gradient ascent is adopted for optimization, thus naturally dealing with issues due to the non-convexity of the solution space.  

We test our method by constructing a portfolio with hundreds of instruments, and considering a testing period of more than 20 years. The experiments show superior performance for our approach,  ensuring the scalability and generalization of the end-to-end framework. We also investigate the model performance under realistic transaction costs, and carry out comparisons to discuss the usage of different deep neural networks. In addition, we compare the classical two-step MV problem with our approach on a simulated data set, where the optimal portfolio weights are known. The results indicate that our method delivers solutions that are closer to optimal weights, showcasing the efficiency of our setup.

\paragraph{Outline} The remainder of this paper is structured as follows: A literature review is included to introduce related works in Section~\ref{literature}. Section~\ref{methods} presents our method and details various constraints. We then describe our experiments in Section~\ref{experiments}, and conclude the findings in Section~\ref{conclusion}, along with future research directions.

\section{Literature Review}
\label{literature}

The portfolio optimization theory, formally introduced by Markowitz \cite{markowitz1952portfolio}, plays a significant role in both research and practice. The basic MV model considers only long positions and requires budget constrains; for example, the sum of weights equals to 1. It can be formulated as a quadratic programming problem \cite{frank1956algorithm}, and the solution only depends on the expected mean and covariance matrix of asset returns. The simplest way of estimating the expected mean and covariance matrix is to use the sample mean and covariance matrix. However, the gain from plugging the sample estimates into the optimization problem is offset by the error between population parameters and their sample estimates, also known as ``Markowitz optimization enigma", revealed by \cite{michaud1989markowitz, lai2011mean, kan2007optimal}.   

To resolve the ``Markowitz optimization enigma", a major direction is to find better estimators of expected means, such as Bayes and shrinkage estimators \cite{jorion1986bayes, lai2011mean, pastor2000comparing}. However, the dynamics of financial returns is rather stochastic and the prediction of returns is a highly non-trivial task. There also exists a large volume of literature applying machine learning models to predict asset returns,  including random forest \cite{ballings2015evaluating}, vanilla neural networks \cite{schoneburg1990stock}, long short-term memory (LSTM) networks \cite{fischer2018deep, fabbri2018dow}, and others. Furthermore, the estimation of the covariance matrix is unstable when we have a large collection of variables \cite{ledoit2003improved, ledoit2004honey}. To remedy this, our framework directly outputs portfolio weights and optimizes objective functions bypassing the estimation of expected means and the covariance matrix of returns. 

Another drawback of the basic MV model is that it neglects financial restrictions that are essential in practice. For example, the cardinality constraint limits the number of assets to be included in a portfolio and the requirement of maximum position for individual instrument is necessary for diversification. In the literature, heuristic mechanisms \cite{holland1992genetic, chang2000heuristics} are proposed to study the cardinality constraint, such as genetic algorithms, simulated annealing, and branch-and-bound algorithms \cite{gao2013optimal, lawler1966branch}. With respect to the maximum position constraint, it can be solved efficiently by some specialized methods such as the simplex method \cite{wolfe1959simplex}. However, the computational complexity of the portfolio optimization problem becomes much greater than the basic model when considering these restrictions. In particular, the cardinality constraint makes the problem fall into the class of NP-hard problems \cite{shaw2008lagrangian, chang2000heuristics}. In our case, we formulate these constraints by utilizing corresponding specialized neural layers, including a differentiable sorting algorithm \cite{grover2018stochastic, cuturi2019differentiable, blondel2020fast}, paving the way for gradient ascent to be used for training, and thus avoiding the computation difficulties.

The idea of combining the prediction and optimization tasks in an end-to-end model has been previously explored in the context of portfolio optimization \cite{zhang2020deep, butler2021integrating, uysal2021end}, as well as in other contexts, such as momentum strategies \cite{DeepMomentum}. Zhang et al.  \cite{zhang2020deep} adopt neural networks to directly optimize the portfolio's Sharpe Ratio by circumventing the prediction of future returns. However, the portfolio in \cite{zhang2020deep} only considers the basic MV problem, i.e. requiring long-only and budget-constrained. Recent work of \cite{butler2021integrating} uses neural networks with differentiable optimization layers to find the solutions of MV problems with certain constraints. The work of \cite{uysal2021end} learns the optimal risk contribution on each asset and allocates the weights according to the risk-budgeting strategy \cite{bruder2012managing}. It is worth to emphasize that all aforementioned portfolios can be optimized and unified in our proposed framework, and we provide a comprehensive study of end-to-end frameworks by optimizing various objective functions with several practical constraints.


\section{Methods}
\label{methods}

In this section, we introduce our end-to-end framework and start with a general mean-variance problem (MVP) that allows for \emph{short-selling}
\begin{equation}
\begin{split}
\max_{\mathbf{w}_{t}} \,\, \mathbb{E}(r_{p, t+1}) - \frac{\lambda}{2} \operatorname{Var}(r_{p, t+1}), &\text{\quad s.t.\quad} \|\mathbf{w}_t\|_1 = \sum_{i=1}^N |w_{i,t}| = 1,\\
r_{p, t+1} &= {\mathbf{w}_t}^{\prime} \mathbf{r}_{t+1},
\end{split}
\label{eq:mvp}
\end{equation}
where $r_{p, t+1}$ is the portfolio return and $\lambda$ is the risk aversion rate. $\mathbf{w}_{t} = (w_{1, t}, \dots, w_{N,t})^{\prime}$ represents the fraction of the portfolio value invested in the each asset at time $t$. $\mathbf{r}_t = (r_{1, t}, \dots, r_{N, t})^{\prime}$ is the collection of returns of $N$ assets at time $t$ and $r_{i, t}$ denotes the percentage return of asset $i  \,\, (i=1, \dots, N)$. Note that $t$ can denote any arbitrary period, such as one-minute intervals, days, months, etc. 


\paragraph{Objective functions.} Besides the MVP defined in Equation~\eqref{eq:mvp}, we study the following objective functions
\begin{itemize}
	\item \textit{Global Minimum Variance Portfolio (GMVP)}
    \begin{equation}
        \min_{\mathbf{w}_t} \, \operatorname{Var}(r_{p, t+1}),
    \end{equation}
	\item \textit{Maximum Sharpe Ratio Portfolio (MSRP)}
	\begin{equation}
        \max_{\mathbf{w}_t} \, \frac{\mathbb{E}(r_{p, t+1})}{\operatorname{Var}(r_{p, t+1})},
    \end{equation} 
    Note that we let all objective functions have the same constraints as in MVP (Equation~\eqref{eq:mvp}), and allow for short-selling. 
\end{itemize}

\paragraph{Constraints.} In terms of the constraints, we consider the following cases
\begin{itemize} \label{constraints}
	\item \textit{Long-only (LONG)}: $w_{i,t} \geq 0$, as short-selling is forbidden in certain markets, for example, in Chinese stock markets,
	\item \textit{Maximum position (MAX)}: $w_{i,t} \leq u$ specifies the upper bound $u$ (with $u>0$) of weights allocated to each asset in a portfolio which helps with concentration risk, ensuring diversification, 
	\item \textit{Cardinality (CAR)}: $\|\mathbf{w}_t\|_0 = K$ restricts the number of assets in a portfolio, 
	\item \textit{Leverage (LEV)}: $\|\mathbf{w}_t\|_1 = L$ with $L \geq 1$ allows investors to increase the exposure to markets.
\end{itemize}





\subsection{The Proposed End-to-End Framework}

\begin{figure}[!t]
    \centering
    \includegraphics[width=.9\textwidth]{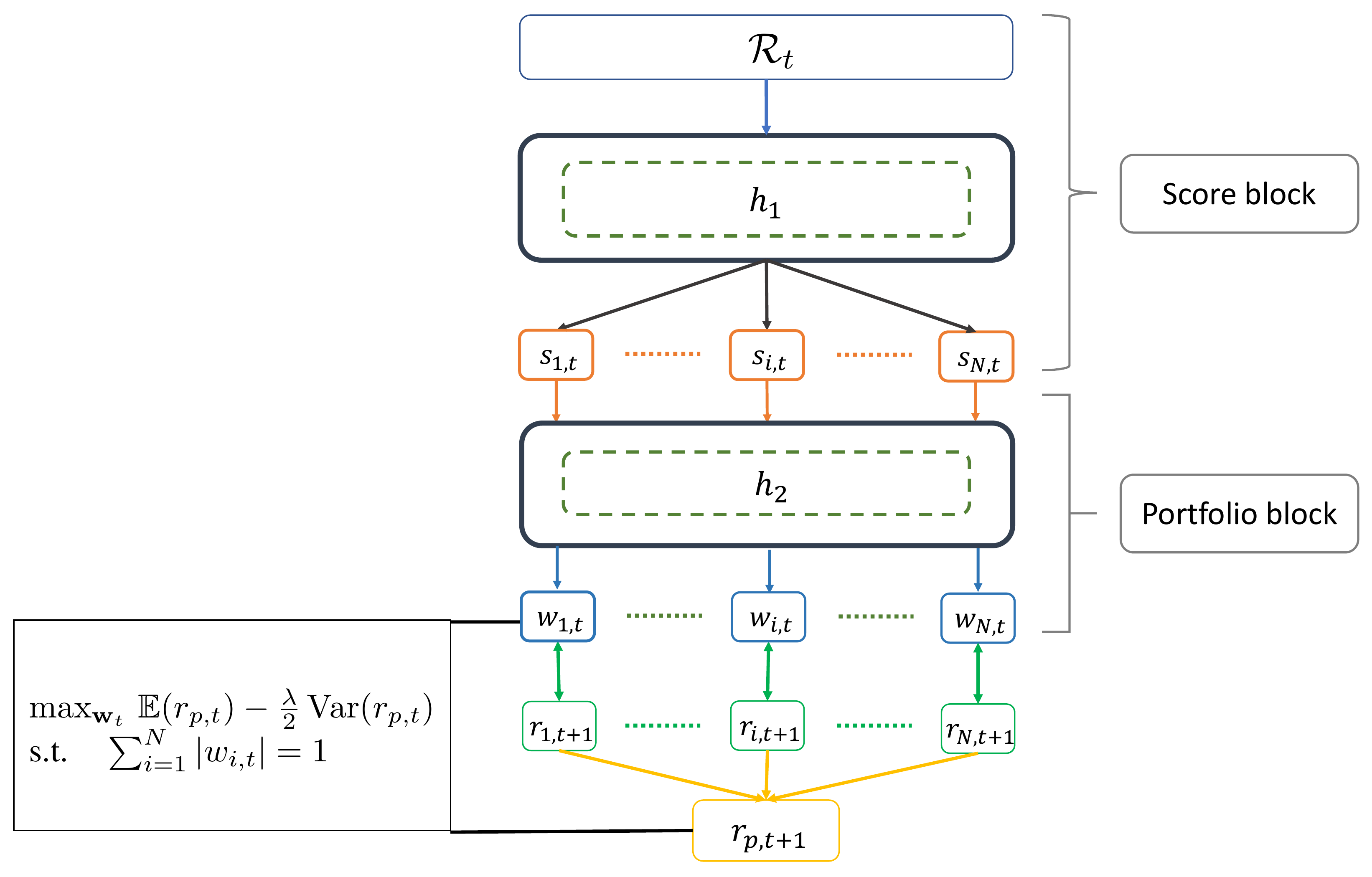}
    \caption{Architecture of our proposed End-to-End approach. $h_1$ represents a neural network, which transforms the input into fitness scores for each asset. $h_2$ represents the specialized differential layers to convert the scores to the portfolio weights satisfying the required constraints.}
    \label{fig:e2e}
\end{figure}

We denote  by  $\mathcal{R}_{t}$   the set of current information, for example, previous returns $\{\mathbf{r}_1, \dots, \mathbf{r}_{t}\}$. The classical approach can be essentially divided into two parts: prediction and optimization. The predictive step is to estimate future returns by $\hat{r}_{i, t+1} = f\left(\mathcal{R}_t\right)$ and the predicted returns are then inserted into the objective function in Equation~\eqref{eq:mvp} to derive the portfolio weights $\mathbf{w}_t$. 
In this work, we directly model the portfolio weights $\mathbf{w}_t$ with deep neural networks ($f$) as
\begin{equation}
\mathbf{w}_t = f\left(\mathcal{R}_t\right),
\end{equation}
where $\mathcal{R}_{t}$ denotes the set of current information as inputs to the networks. Figure \ref{fig:e2e} depicts our proposed end-to-end framework, which is composed of two ingredients: a score block and a portfolio block.

\paragraph{Score block.} The input $\mathcal{R}_t$ consists of current market information, for example, returns $\mathbf{r}_{t-p}, \dots, \mathbf{r}_{t}$ up to lag $p$. We use a neural network to transform the input into fitness scores for each asset. A higher score implies that the corresponding asset is more likely to be assigned higher weights (without cardinality constraints) or selected (with cardinality constraints), and vice versa. We denote this network function as $h_1$ and the fitness scores as
\begin{equation}
    \mathbf{s}_t = (s_{1, t}, \dots, s_{N, t})^{\prime} = h_1(\mathcal{R}_t).
\end{equation}

\paragraph{Portfolio block.} In the portfolio block, we convert the previous obtained scores $\mathbf{s}_t$ into the portfolio weights satisfying the required constraints through corresponding differential functions $h_2(\cdot)$. We then compute the realized portfolio returns $r_{p, t+1}$ based on future returns $\mathbf{r}_{t+1}$ and calculate the loss value according to the selected objective function, for example, Equation~\eqref{eq:mvp}. To allow for short selling and to meet the requirement of the absolute values of the weights to sum to one, we transform the fitness scores as follows,
\begin{equation} \label{eq:short_selling}
    \begin{split}
        w_{i,t} &= h_2 (s_{i,t}) \\
        &= \text{sign} (s_{i,t}) \times \frac{e^{s_{i, t}}}{\sum_{j=1}^{N} e^{s_{j, t}}}.
    \end{split}
\end{equation}


\subsection{The Design of the Differential Function $h_2$ for Constraints}

Similar to Equation~\eqref{eq:short_selling}, we now demonstrate how the differential function $h_2$ can be constructed in order to meet the previous discussed constraints, by considering the fitness scores $\bm{s}_t$ from the score block. We first showcase how a single constraint can be incorporated by 
re-parameterizing the scores, and then demonstrate how multiple constraints can be combined.

\paragraph{Re-parametrisation of $\mathbf{s}_t$ for a single constraint} 
\

\noindent \textbf{Constraint (1)} \textit{Long-only} and $\|\mathbf{w}_t\|_1 = 1$: We apply the  $\operatorname{softmax}$  activation function to the scores, in order to obtain the non-negative weights. For $i=1, \ldots, N$, we consider
\begin{equation}
    w_{i, t} = \frac{e^{s_{i, t}}}{\sum_{j=1}^{N} e^{s_{j, t}}}.
\end{equation}

\noindent \textbf{Constraint (2)} \textit{Maximum} and $\|\mathbf{w}_t\|_1 = 1$: We use a generalized sigmoid function, $\phi_a(x) = a + \frac{1}{1+e^{-x}}$ (with $a \geq 0$) to transform the scores, so that weights can automatically meet the upper bound constraint $u$. Upon setting $a = \frac{1-u}{Nu-1}$, we obtain $w_{i,t} \leq u$ as follows 
\begin{equation}
w_{i, t} = \operatorname{sign}(s_{i, t}) \times \frac{\phi_a(|s_{i, t}|)}{\sum_{j=1}^{N} \phi_a(|s_{j, t}|)}.
\end{equation}
If we let the maximum position $u$ equal to 1, $\phi_a(x)$ degenerates to the sigmoid function.

\noindent \textbf{Constraint (3)} \textit{Cardinality} and $\|\mathbf{w}_t\|_1 = 1$: 
To achieve cardinality, we first employ a sorting operator $\Pi(\cdot)$ that maps $\mathbf{s}_t \in \mathbb{R}^{N}$ into a permutation matrix $\Pi(\mathbf{s}_t) \in \mathbb{R}^{N\times N}$, such that $\tilde{\mathbf{s}}_t = \Pi(\mathbf{s}_t)$ is the sorted vector of $\mathbf{s}_t$ in descending order.
We then long the top $n$ instruments and short the bottom $n$ instruments to construct our portfolio
\begin{align}
\begin{split}
w_{i, t} &= \frac{1}{2} \times \frac{{1}_{\{{s}_{i, t} > d_{u}\}} e^{|s_{i, t}|}}{\sum_{j=1}^{N} {1}_{\{{s}_{j, t} > d_{u}\}} e^{|s_{j, t}|}} - \frac{1}{2} \times \frac{{1}_{\{{s}_{j, t} < d_{l}\}} e^{|s_{i, t}|}}{\sum_{j=1}^{N} {1}_{\{{s}_{i, t} < d_{l}\}} e^{|s_{j, t}|}}, \\
d_u &= \tilde{\mathbf{s}}_t[n], \quad d_{l} = \tilde{\mathbf{s}}_t[N-n], \quad n = \floor{K/2} + 1, \\
\end{split}
\label{eqn:cardinality}
\end{align}
where $\tilde{\mathbf{s}}_t[n]$ means the $n$-th element in vector $\tilde{\mathbf{s}}_t$, i.e. the $n$-th largest element of $\mathbf{s}_t$.
To compute the sorting operator, we first define a square matrix $\Lambda_{i,j}^{t}$ on the fitness score $\mathbf{s}_t$ as
\begin{equation}
    \Lambda_{i,j}^{t} = (N+1-2i) s_{j,t} - \sum_{m} |s_{j,t} - s_{m,t}|.
\end{equation}
\vspace{-2mm}

According to previous works \cite{ogryczak2003minimizing, grover2018stochastic, blondel2020fast, cuturi2019differentiable}, the permutation matrix $\Pi(\mathbf{s}_t)$ can be constructed as 
\vspace{-1mm}
\begin{equation}
  \Pi(\mathbf{s}_t)_{i,j} =
    \begin{cases}
      1, & \text{if}\  j = \operatorname{argmax} (\Lambda_{i,:}^t) \\
      0, & \text{otherwise}.
    \end{cases}       
\end{equation}

Since  the $\operatorname{argmax}$ function is not differentiable, the authors of \cite{grover2018stochastic} proposes a NeuralSort layer that replaces the $\operatorname{argmax}$  operator by the $\operatorname{softmax}$ operator, to arrive at a relaxed version of $\Pi(\mathbf{s}_t)$
\begin{equation}
    \widehat{\Pi(\mathbf{s}_t)}_{i,:} = \operatorname{softmax}(\Lambda_{i,:}^t).
\end{equation}
Then Equation \eqref{eqn:cardinality} is differentiable and the standard gradient descent method can be applied.

\noindent \textbf{Constraint (4)} \textit{Leverage, i.e. $\|\mathbf{w}_t\|_1 = L$}: Similar to Equation~\eqref{eq:short_selling}, we increase the total  exposure of the positions by a factor of $L$
\begin{equation}
    w_{i, t} = L \times  \operatorname{sign}(s_{i, t}) \times \frac{e^{|s_{i, t}|}}{\sum_{j=1}^{N} e^{|s_{j, t}|}}.
\end{equation}

\paragraph{Re-parametrisation of $\mathbf{s}_t$ for multiple constraints.} 
We handle multiple constraints by combining the corresponding techniques in the above cases. Consider the constraints \textit{Maximum \& Cardinality \& Leverage} as an example, using the techniques in the Cases (2), (3) and (4), we have, for $i=1, \ldots, N$,
\begin{align}
\begin{split}
w_{i, t} &= \frac{L}{2} \times \frac{{1}_{\{{s}_{i, t} > d_{u}\}} \phi_a(|s_{i, t}|)}{\sum_{j=1}^{N} {1}_{\{{s}_{j, t} > d_{u}\}} \phi_a(|s_{j, t}|)} - \frac{L}{2} \times \frac{{1}_{\{{s}_{i, t} < d_{l}\}} \phi_a(|s_{i, t}|)}{\sum_{j=1}^{N} {1}_{\{{s}_{j, t} < d_{l}\}} \phi_a(|s_{j, t}|)}, \\
d_u &= \tilde{\mathbf{s}}_t[n], \quad d_{l} = \tilde{\mathbf{s}}_t[N-n], \quad n = \floor{K/2} + 1, \\
\end{split}
\end{align}
where $a = \frac{1-u/2L}{nu/2L-1}$, and one may analogously perform the re-parametrisation for other combinations of constraints.

As an example, we can consider a long-short equity portfolio.
This is a popular strategy where we go long a certain percentile of top performing  stocks, and short the same percentile of bottom stocks. Commonly, in the literature, one first predicts the assets' returns and sorts assets in order. Interestingly, this strategy can be achieved through our framework by incorporating the third  \textit{Cardinality} constraint. As fitness scores are sorted from the \emph{Score block} in Figure~\ref{fig:e2e}, we can simply long the top best assets and short the bottom worst ones. This also provides an alternative to recently proposed learning-to-rank algorithms to construct such long-short strategies \cite{LearningToRank}.


\section{Experiments}
\label{experiments}

\subsection{Descriptions of Data sets}

In this section, we demonstrate the efficacy of our methods on two data sets. The first one is a synthetic data set generated from multivariate normal distributions, where we compare the classical two-step method with our end-to-end framework, in terms of the similarity of the optimal weights. The second data set contains daily observations for 735 stocks from the Wharton Research Data Services (WRDS) \cite{wrds}, and we test the generalization of our model with a testing period of more than 20 years.

\paragraph{Synthetic data set.}
We simulate daily returns from multivariate normal distributions with parameters calibrated from real data. Specifically, in terms of the mean and covariance matrix, we use the sample mean and (shrunk) sample covariance matrix corresponding to each year from 1984 to 2020 of the WRDS data. 

\paragraph{The WRDS data set} 
We extract daily observations for 735 stocks from the Russell 3000 Index, ranging from 1984/01/03 to 2021/07/06. A rolling window approach is adopted for training; in our case, we consider data between 1984 and 1999 as the first training set, 2000 as the validation set,  and 2001 as the testing set. We then roll forward one year and repeat the process until the end date, gradually increasing the training size. Overall, our testing period ranges from 2001 to 2021.

\subsection{Training and Testing Procedure}
\label{train_test_procedure}
We denote our end-to-end framework as \textsc{E2E}  and compare it with the classical setup (CS). Several models are tested in the experiments, including a linear model (LM), a single fully connected neural network layer with 64 units, also referred to as multilayer perceptron (MLP), a single LSTM layer with 64 units and a convolutional neural network (CNN) model. The CNN model consists of 4 layers where the first 3 layers are one-dimensional convolutional layers with filters of size 32, 64, 128, and each filter has the same kernel size (3,1). The last layer of the CNN model is a single LSTM with 64 units. We set the learning rate to $10^{-4}$, batch size to $64$ and number of epochs to $1000$.
 
When reporting performance during the testing period, we include transaction costs and calculate the portfolio return as 
\begin{equation}
\begin{split}
R_{p,t} &= \sum_{i=1}^{N} R_{i,t} \\ 
R_{i,t} &= w_{i,t-1} r_{i,t} - C |w_{i,t} (1+w_{i,t-1}r_{i,t}) - w_{i,t-1} |,
\end{split}
\end{equation}
where $R_{i,t}$ is the trade return for the instrument $i$ and $r_{i,t}$ its price return at time $t$. We denote the cost rate, $C$, in terms of basis points ($1 bp = 10^{-4}$) and report the following metrics to evaluate model performance
\begin{itemize}
    \item E(R): annualized expected return,
    \item Std(R): annualized standard deviation,
    \item Sharpe: annualized Sharpe Ratio,
    \item DD(R): annualized downside deviation, where only the standard deviation of negative returns is calculated,
    \item Sortino: annualized Sortino Ratio,
    \item MDD: maximum drawdown,
    \item \% of +Ret: the ratio between positive and negative returns,  
    \item Frobenius norm: the similarity between the estimated portfolio weights and the optimal weights. The smaller the value, the better the estimation.
    \item Turnover: an indicator that represents the turnover rate,
    \item Beta: it measures a portfolio's correlation relative to S\&P 500 Index.
\end{itemize}


\subsection{Experimental Results for the Synthetic Data Set}
\label{synthetic_dataset}

In Table~\ref{tab:results_simulation}, we present the results for the synthetic data set when maximizing the Sharpe Ratio, with the  constraint requiring that the summation of the absolute values of the weights equals to one. We denote the classical two-step method as CS, and our end-to-end approach as \textsc{E2E}. Each framework is tested by four models described in Section~\ref{train_test_procedure}, and we include a naive  estimation for CS by calculating the predicted return with just past returns (CS-SAMPLE).  

We focus on Sharpe Ratio and Frobenius norm as the main metrics. The Sharpe Ratio essentially describes the return per unit of risk and our \textsc{E2E} approach demonstrates superior results compared to the CS framework. Additionally, all \textsc{E2E} models deliver better portfolio weights in terms of Frobenius norm, meaning that the weights derived from the end-to-end setup are closer to the optimal weights, when compared to the classical predictive approach.

\begin{table}[!t]
    \centering
    \resizebox{1.0\textwidth}{!}{\begin{tabular}{l|lllllllll}
\toprule
                     & \textbf{E(R)}   &  \textbf{Std(R)} & \textbf{Sharpe} & \textbf{DD(R)}  & \textbf{Sortino} & \textbf{MDD}   & \textbf{$\frac{\text{\% of}}{\text{+Ret}}$} & \textbf{$\frac{\text{Frobenius}}{\text{norm}}$} &\textbf{Turnover}\\
\midrule
\multicolumn{10}{l}{\textbf{MSRP}}               \\
\midrule
CS-SAMPLE   & 0.011  & 0.018  & 0.639  & 0.012    & 0.932   & 0.072 & 0.517   & 4.695   &2.006  \\
CS-LM    & 0.001  & 0.011  & 0.087  & 0.008    & 0.123   & 0.037 & 0.501   & 4.700   &9.635  \\
CS-MLP    & 0.001  & 0.017  & 0.011  & 0.012    & 0.016   & 0.082 & 0.497   & 4.873   &9.518  \\
CS-LSTM   & 0.001  & 0.011  & 0.053  & 0.008    & 0.074   & 0.086 & 0.504   & 4.684   &6.913  \\
CS-CNN    & 0.002  & 0.011  & 0.160  & 0.008    & 0.227   & 0.053 & 0.500   & 4.685   &7.214  \\
E2E-LM    & 0.004  & 0.031  & 0.146  & 0.022    & 0.206   & 0.119 & 0.513   & \textbf{4.253}   &7.203  \\
E2E-MLP    & 0.003  & 0.012  & 0.238  & 0.009    & 0.339   & 0.056 & 0.516   & 4.280   &5.849  \\
E2E-LSTM    & 0.014  & 0.012  & \textbf{1.216}  & 0.008    & \textbf{1.810}   & 0.008 & 0.529   & 4.407   &0.692  \\
E2E-CNN    & 0.014  & 0.014  & 0.977  & 0.009    & 1.441   & 0.027 & 0.534   & 4.426   &3.604  \\
\bottomrule
\end{tabular}}
	\caption{Experimental results for the synthetic data set.}
    \label{tab:results_simulation}
\end{table}

\subsection{Experimental Results for the WRDS Data Set}

Table~\ref{tab:results_real} shows the experimental results for the WRDS data set, with a testing period of more than 20 years ranging from 2001 to 2021. The table is split into four parts, where the first part (Baselines) includes five benchmark models: the performance of the S\&P 500 Index, a portfolio with equal weights (EWP), the maximum diversification (MD) portfolio \cite{theron2018maximum}, the diversity-weighted portfolio (DWP) from stochastic portfolio theory \cite{kom2016stochastic} and the global minimum variance portfolio (GMVP) \cite{theron2018maximum}. The second block of Table~\ref{tab:results_real} compares our approach with the classical predictive methods. The objective function is the same as in the previous Section~\ref{synthetic_dataset}, where we maximize the Sharpe Ratio. The third block studies various other objective functions, and the last block presents the effects of different constraints. In particular, we demonstrate the following cases
\begin{itemize}
    \item MSRP-LONG: a long-only portfolio that maximizes the Sharpe Ratio,
    \item LEV: a portfolio with leverage, where the leverage is set to 5,
    \item CAR: a portfolio with a cardinality constraint set to 148, which amounts to 20\% of our total number of instruments; in other words, we long the top decile and short the bottom decile, 
    \item MAX: a portfolio with a maximum position for each individual instrument, which we set to 5\%, 
    \item LEV-CAR-MAX: a combination of above constraints.
\end{itemize}

We draw the following conclusions. From the second block of Table~\ref{tab:results_real}, we observe that our end-to-end methods deliver better results than the classical approach and outperform all the baseline models. The third block demonstrates how different objective functions affect the model performances. Specifically, GMVP attains the lowest variance, and we can achieve the desired risk level by adjusting the risk aversion parameter ($\lambda$) in MVP, where a higher $\lambda$ penalizes the risk more, leading to a portfolio with a smaller variance. The last block shows the results for different constraints, and we include them as examples to showcase the  flexibility of our method. The exact parameters of these constraints can be
customized by  
the end-user, in order to achieve their preferred conditions and adjust to the specificity of their trading environment.  Table~\ref{tab:results_real_cost} has the same layout as Table~\ref{tab:results_real}, but we study model performances under transaction costs ($C=2$ bps). 

\begin{table}[!t]
    \centering
    \resizebox{1.025\textwidth}{!}{\begin{tabular}{l|lllllllll}
\toprule
                     & \textbf{E(R)}   &  \textbf{Std(R)} & \textbf{Sharpe} & \textbf{DD(R)}  & \textbf{Sortino} & \textbf{MDD}   & \textbf{$\frac{\text{\% of}}{\text{+Ret}}$} & \textbf{Beta} &\textbf{Turnover}\\
\midrule
\multicolumn{10}{l}{\textbf{Baselines}}               \\
\midrule
S\&P 500   & 0.061  & 0.196  & 0.402  & 0.140    & 0.563   & 0.568 & 0.541   & 1.000   &0.000  \\
EWP    & 0.130  & 0.212  & 0.682  & 0.148    & 0.973   & 0.548 & 0.546   & 1.000   &0.000  \\
MD    & 0.439  & 0.239  & 1.641  & 0.141    & 2.785   & 0.519 & 0.548   & 0.599   &2.347  \\
DWP    & 0.117  & 0.209  & 0.633  & 0.147    & 0.899   & 0.554 & 0.545   & 1.000   &0.009  \\
GMVP    & 0.080  & 0.081  & 0.992  & 0.059    & 1.360   & 0.408 & 0.564   & 0.257   &1.577  \\
\midrule
\multicolumn{10}{l}{\textbf{MSRP}}               \\
\midrule
CS-SAMPLE   & -0.031  & 0.038  & -0.812  & 0.028    & -1.093   & 0.476 & 0.484   & 0.037   &1.708  \\
CS-LM    & 0.004  & 0.015  & 0.290  & 0.011    & 0.414   & 0.062 & 0.504   & 0.009   &9.157  \\
CS-MLP    & 0.008  & 0.027  & 0.299  & 0.019    & 0.424   & 0.140 & 0.515   & 0.036   &9.319  \\
CS-LSTM   & 0.014  & 0.017  & 0.858  & 0.011    & 1.259   & 0.043 & 0.520   & 0.014   &7.624  \\
CS-CNN    & 0.007  & 0.017  & 0.426  & 0.012    & 0.609   & 0.093 & 0.513   & 0.014   &4.425  \\
E2E-LM    & 0.049  & 0.044  & 1.116  & 0.030    & 1.649   & 0.168 & 0.546   & 0.011   &7.388  \\
E2E-MLP    & 0.044  & 0.026  & 1.688  & 0.016    & 2.657   & 0.073 & 0.552   & \textbf{0.008}   &7.124  \\
E2E-LSTM    & 0.060  & 0.023  & \textbf{2.604}  & 0.013    & \textbf{4.448}   & 0.017 & 0.578   & 0.017   &2.840  \\
E2E-CNN    & 0.023  & 0.024  & 0.931  & 0.017    & 1.365   & 0.084 & 0.531   & 0.046   &3.881  \\
\midrule
\multicolumn{10}{l}{\textbf{Other Objective Functions}}               \\
\midrule
LSTM-GMVP    & 0.001  & \textbf{0.011}  & 0.047  & 0.008    & 0.067   & 0.060 & 0.504   & -0.004   &0.170  \\
LSTM-MVP$_{\lambda=10}$  & 0.064  & 0.317  & 0.349  & 0.207    & 0.534   & 0.878 & 0.518   & 0.342   &4.209  \\
LSTM-MVP$_{\lambda=20}$  & 0.179  & 0.169  & 1.055  & 0.115    & 1.555   & 0.380 & 0.548   & 0.195   &3.910  \\
LSTM-MVP$_{\lambda=30}$  & 0.168  & 0.116  & 1.394  & 0.076    & 2.149   & 0.187 & 0.552   & 0.060   &3.483  \\
\midrule
\multicolumn{10}{l}{\textbf{Constraints}}               \\
\midrule
LSTM-MSRP-LONG    & 0.368  & 0.197  & 1.691  & 0.125    & 2.666   & 0.253 & 0.566   & 0.767   &5.309  \\
LSTM-MSRP-LEV    & 0.321  & 0.112  & 2.540  & 0.068    & 4.203   & 0.151 & 0.576   & 0.132   &13.29  \\
LSTM-MSRP-CAR    & 0.032  & 0.056  & 0.588  & 0.039    & 0.844  & 0.167 & 0.520   & -0.011   &8.638  \\
LSTM-MSRP-MAX    &0.057  & 0.021  & \textbf{2.683}  &0.012    & \textbf{4.459}   & 0.026 & 0.578   & \textbf{0.021}   &2.671  \\
LSTM-MSRP-LEV-CAR-MAX    & 0.121  & 0.140  & 0.885  & 0.093    & 1.340   & 0.296 & 0.518   & 0.029   &18.49  \\
\bottomrule
\end{tabular}}
	\caption{Experimental results for the WRDS dataset with C = 0 bp.}
    \label{tab:results_real}
\end{table}

\begin{table}[!t]
    \centering
    \resizebox{1.025\textwidth}{!}{\begin{tabular}{l|lllllllll}
\toprule
                     & \textbf{E(R)}   &  \textbf{Std(R)} & \textbf{Sharpe} & \textbf{DD(R)}  & \textbf{Sortino} & \textbf{MDD}   & \textbf{$\frac{\text{\% of}}{\text{+Ret}}$} & \textbf{Beta} &\textbf{Turnover}\\
\midrule
\multicolumn{10}{l}{\textbf{Baselines}}               \\
\midrule
S\&P 500   & 0.061  & 0.196  & 0.402  & 0.140    & 0.563   & 0.568 & 0.541   & 1.000   &0.000  \\
EWP    & 0.129  & 0.212  & 0.678  & 0.148    & 0.968   & 0.549 & 0.546   & 1.000   &0.000  \\
MD    & 0.414  & 0.239  & 1.569  & 0.141    & 2.649   & 0.531 & 0.546   & 0.599   &2.347  \\
DWP    & 0.116  & 0.209  & 0.630  & 0.147    & 0.895   & 0.555 & 0.545   & 1.000   &0.009  \\
GMVP    & 0.068  & 0.081  & 0.849  & 0.060    & 1.156  & 0.421 & 0.562   & 0.257   &1.577  \\
\midrule
\multicolumn{10}{l}{\textbf{MSRP}}               \\
\midrule
CS-SAMPLE   & -0.043  & 0.038  & -1.136  & 0.029    & -1.505   & 0.593 & 0.473   & 0.037   &1.708  \\
CS-LM    & -0.060  &0.015  & -4.082  & 0.013    & -4.777   & 0.719 & 0.382   & 0.008   &9.160  \\
CS-MLP    & -0.058  & 0.027  & -2.175  & 0.021    & -2.760   & 0.710 & 0.440   & 0.036   &9.319  \\
CS-LSTM   & -0.040  & 0.017  & -2.439  & 0.013    & -3.065   & 0.573 & 0.424   & 0.014   &7.624  \\
CS-CNN    & -0.025  & 0.017  & -1.486  & 0.013    & -1.935   & 0.407 & 0.461   & 0.014   &4.425  \\
E2E-LM    & -0.005  & 0.044  & -0.090  & 0.031    & -0.127   & 0.257 & 0.487   & 0.011   &7.388  \\
E2E-MLP    & -0.008  & 0.026  & -0.295  & 0.018    & -0.424   & 0.223 & 0.477   & \textbf{0.007}   &7.124  \\
E2E-LSTM    & 0.039  & 0.023  & \textbf{1.693}  & 0.014    & \textbf{2.743}   & 0.019 & 0.548  & 0.017   &2.840  \\
E2E-CNN    & -0.006  & 0.024  & -0.217  & 0.017    & -0.302   & 0.261 & 0.487   & 0.046   &3.881  \\
\midrule
\multicolumn{10}{l}{\textbf{Other Objective Functions}}               \\
\midrule
LSTM-GMVP    & -0.001  & 0.011  & -0.105  & 0.008    & -0.147   & 0.070 & 0.499   & -0.004   &0.170  \\
LSTM-MVP$_{\lambda=10}$  & 0.032  & 0.317  & 0.253  & 0.208    & 0.385   & 0.917 & 0.512   & 0.342   &4.209  \\
LSTM-MVP$_{\lambda=20}$  & 0.146  & 0.170  & 0.888  & 0.116    & 1.296   & 0.422 & 0.542   & 0.195   &3.910  \\
LSTM-MVP$_{\lambda=30}$  & 0.139  & 0.117  & 1.177  & 0.077    & 1.792   & 0.238 & 0.542   & 0.060   &3.483  \\
\midrule
\multicolumn{10}{l}{\textbf{Constraints}}               \\
\midrule
LSTM-MSRP-LONG    & \textbf{0.317}  & 0.199  & 1.498  & 0.126    & 2.337   & 0.254 & 0.558   & 0.767   &5.309  \\
LSTM-MSRP-LEV     & 0.200  & 0.113  & 1.679  & 0.071    & 2.647   & 0.169 & 0.548   & 0.131  &13.29  \\
LSTM-MSRP-CAR    & -0.030  & 0.056  & -0.529  & 0.041    & -0.722   & 0.564 & 0.480   & -0.011   &8.638\\
LSTM-MSRP-MAX    & 0.037  & 0.021  & \textbf{1.738}  & 0.013    & \textbf{2.741}   & 0.030 & 0.551   & \textbf{0.021}   &2.671  \\
LSTM-MSRP-LEV-CAR-MAX     & -0.019  & 0.140  & -0.063  & 0.097    & -0.091   & 0.618 & 0.490  & 0.029   &18.49  \\
\bottomrule
\end{tabular}}
	\caption{Experimental results for the WRDS dataset with C = 2 bps.}
    \label{tab:results_real_cost}
\end{table}

\begin{figure}[p]
\centering
\includegraphics[width=1 \textwidth]{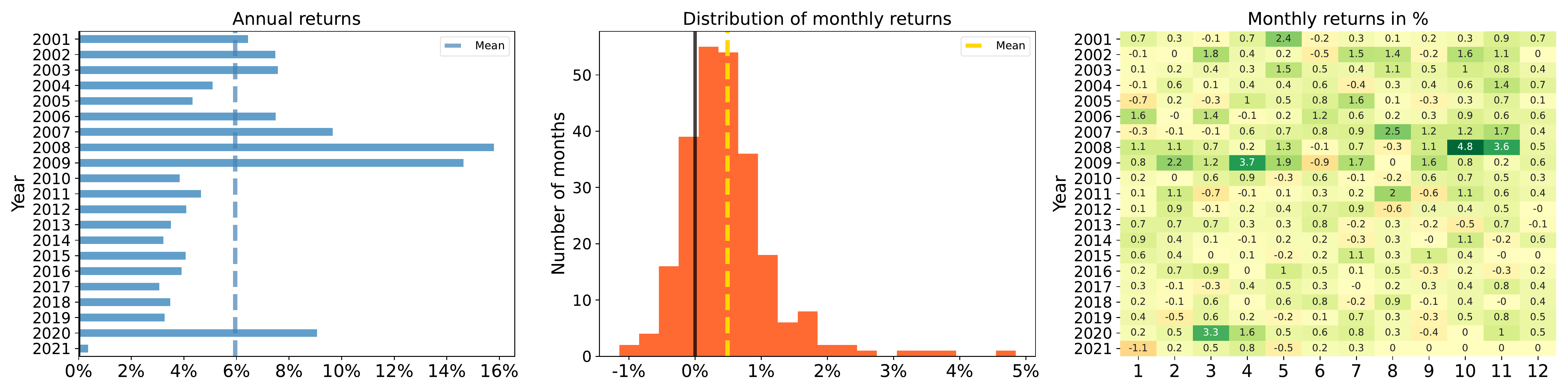}
\includegraphics[width=1 \textwidth]{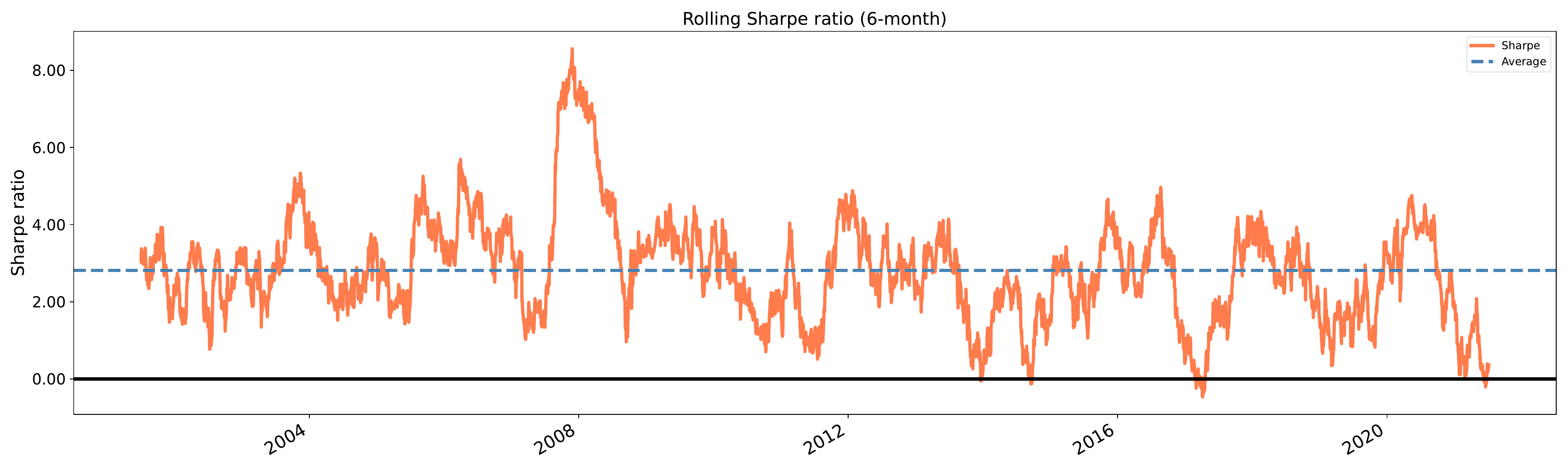}
\includegraphics[width=1 \textwidth]{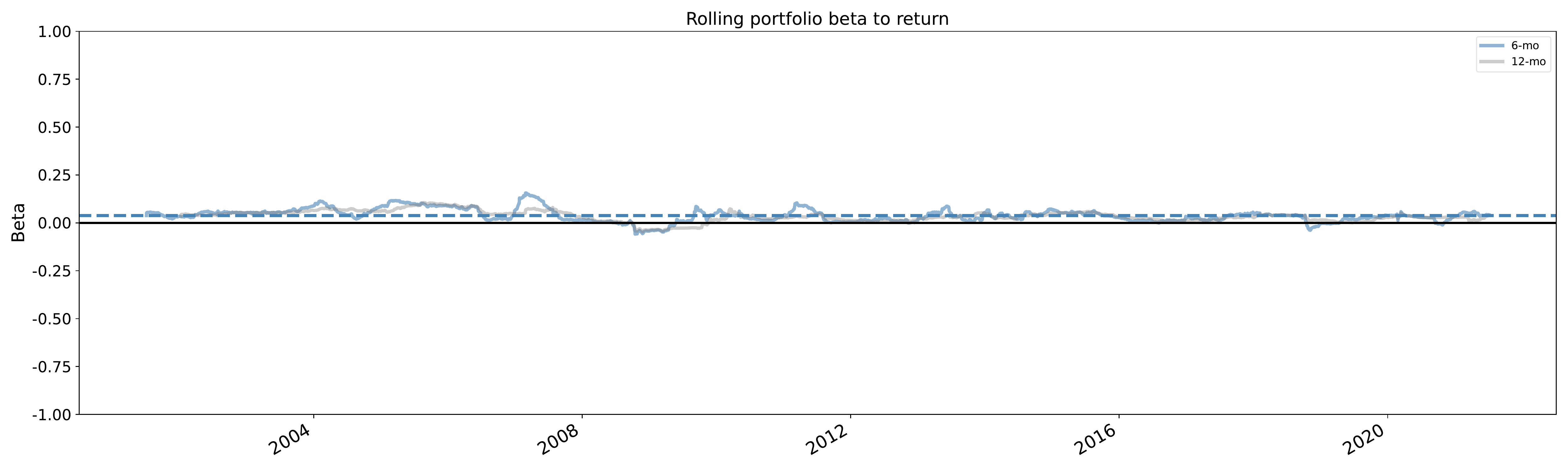}
\includegraphics[width=1 \textwidth]{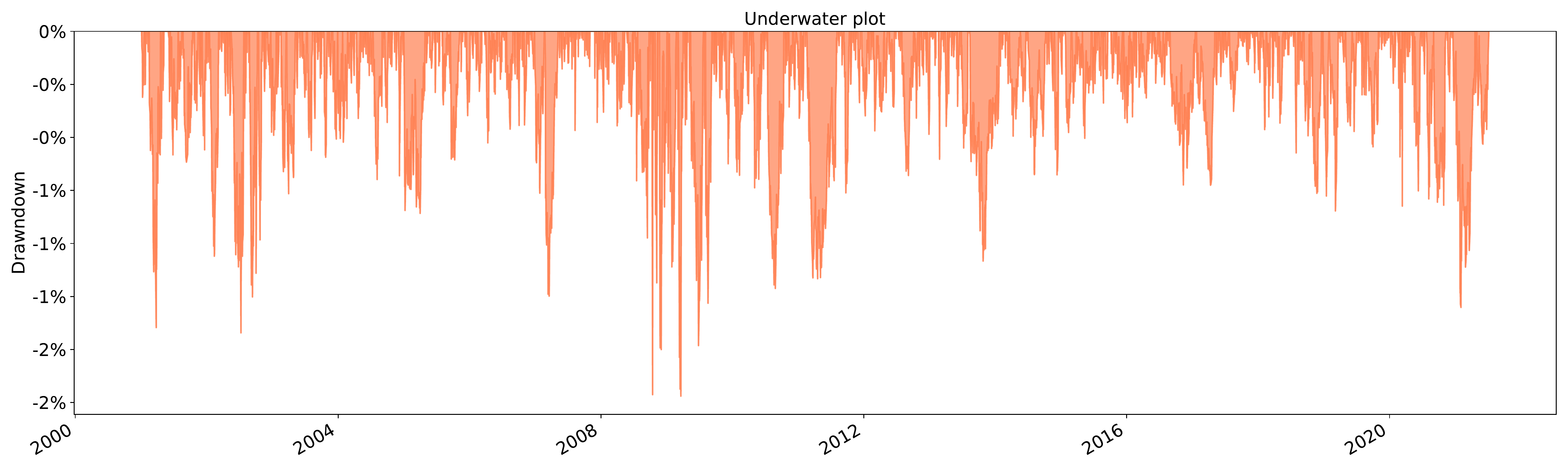}
\caption{Trading performance for E2E-LSTM. \textbf{Top:} Cumulative returns; \textbf{Top Middle:} Rolling Sharpe ratio over time; \textbf{Middle Bottom:} Rolling beta over time; \textbf{Bottom:} Drawdown plot.}
\label{fig:e2e_lstm}
\end{figure}

To assess the robustness of our results, we look at the performance of E2E-LSTM across the entire testing period. Figure~\ref{fig:e2e_lstm} shows the cumulative returns, rolling Sharpe Ratio, rolling beta against S\&P 500 and a drawdown plot for E2E-LSTM ($C=0$ bps). In general, we obtain decent monthly returns without consistent losses, and the rolling Sharpe Ratio is mostly above zero. Since our method allows for short selling, the rolling beta is always close to zero indicating a market-neural strategy; therefore, our model is agnostic to market movements. This observation can be also verified in the drawdown plot -- we observe that drawdown is well controlled during the 2008 crisis and 2020 market crash due to the pandemic, demonstrating robustness of the approach even throughout turbulent periods.

\section{Conclusion}
\label{conclusion}

In this work we propose a universal end-to-end framework for portfolio optimization where portfolio weights are directly computed via a deep learning architecture. 

Our proposed  pipeline  bypasses the traditional  two-stage procedure of first forecasting returns and then computing portfolio weights based on the inverse of the estimated covariance matrix. It thus entirely avoids the challenging problem of estimating the asset covariance matrix. Within our deep learning framework we have the flexibility of optimizing various objective functions including the Sharpe Ratio, mean-variance trade-off and others. Furthermore, several neural network layers are designed to impose constraints on the portfolio weights, such as cardinality, maximum position and leverage. These constraints are popular in practice but suffer from computational complexity when using classical approaches as commonly discussed in the literature. We  
incorporate these constraints into the objective functions through special neural network layers and optimize the entire framework via gradient ascent. 

We test the efficacy of our methods on two data sets. The first one is a synthetic data set where we compare the estimated weights from the classical predictive approach and our end-to-end framework against the optimal weights which are known from the data generating process. The results showcase the accuracy of our method, and demonstrates that the weights derived by our approach are ``closer'' to the optimal weights. 
The second data set consists of hundreds of stocks from WRDS, with a testing period of more than 20 years. Our model delivers superior results in terms of Sharpe Ratio, and we demonstrate how to use constraints to achieve the desired portfolio, for example, using leverage to increase returns or imposing maximum position for individual instrument to prevent concentration of risk.

In subsequent continuations of this work, we aim to integrate transaction costs into the objective functions in order for the turnover of a portfolio to be controlled during training process; we expect the resulting method to be able to tolerate much higher cost rates. This idea is closely related to Reinforcement Learning (RL) \cite{sutton2018reinforcement}, and future extensions of our framework will incorporate insights from the RL literature.

\section*{Acknowledgements}
The authors would like to thank Oxford-Man Institute of Quantitative Finance for computation support and data access.




\end{document}